\documentclass[1p]{elsarticle}
\usepackage[OT4]{fontenc}
\usepackage{hyperref}
\usepackage{graphicx}
\usepackage{psfrag}
\usepackage{subfig}
\usepackage{amsmath}

\begin{document}

\begin{frontmatter}

\title{Mean field model of a game for power}

\author[nasu]{Tatiana Karataieva}
\author[nasu]{Volodymyr Koshmanenko}
\author[agh1] {Ma{\l}gorzata J. Krawczyk}
\author[agh1]{Krzysztof~Ku{\l}akowski\corref{cor}}
\ead{kulakowski@fis.agh.edu.pl}

\address[nasu]{ Institute of Mathematics of NASU, 3 Tereshchenkivska St., 01601, Kyiv, Ukraine
}

\address[agh1]{AGH University of Science and Technology,
Faculty of Physics and Applied Computer Science,
al.~Mickiewicza~30, 30-059 Krak\'ow, Poland.
}

\cortext[cor]{Corresponding author}

%\date{\today}

\begin{abstract}
Our aim is to model a game for power as a dynamical process, where an excess of power possessed by a player allows him
to gain even more power. Such a positive feedback is often termed as the Matthew effect. Analytical and numerical methods
allow to identify a set of fixed points of the model dynamics. The positions of the unstable fixed points give an insight
on the basins of attraction of the stable fixed points. The results are interpreted in terms of modeling of coercive power.
\end{abstract}

%\keyword s{Athermal phase transition; Crowd dynamics; Long-range interactions; Computer simulation}

\end{frontmatter}

%% ===========================================================================
\section{Introduction}
%% ===========================================================================

In his famous essay \cite{1}, Robert Merton has introduced the so-called Matthew effect to social sciences,
when discussing biased distribution of recognition for scientific achievements. As Merton puts it:
'... the Matthew effect consists in the accruing of greater increments of recogition (...) to scientists
of considerable repute and the withholding of such recognition from scientists who have not yet made
their mark.' Since then, the Matthew effect has been the subject of research in education \cite{2}, technology \cite{3},
economy \cite{3,3a,3b}, statistics \cite{4}, and science again \cite{5,6,7}, to name only a few \cite{3,8,8a,8b}. The effect, commonly
cited as 'accumulated advantage' or 'rich gets richer', can be defined as a positive feedback between an
amount of possessed goods and an ability of acquiring even more goods. Even when the term 'Matthew effect'
is not cited literally, the phenomenon itself is at the centre of attention of historians, sociologists,
economists and political scientists \cite{9,10,10a,10b,10c}. \\

Our interest is modeling of dynamics of power, one of central concept in sociology \cite{astwr}. We accept the
classical definition by Max Weber: power is 'the ability of an individual or group to achieve their own goals or aims
when others are trying to prevent them' \cite{mweb}. More specifically, we imagine a zero-sum game for coercive power,
with the latter not based on a social structure but rather on individual characteristics of social actors.
In social simulations, the Matthew effect is often called to interpret the assumption of preferential
attachment in growing networks \cite{11,11a}. In this sense 'rich gets richer' means that a node (actor) of
large degree (number of neighbors) has more chances to get even more neighbors, and therefore her/his
position in the network, as measured by centrality, betweenness etc. \cite{11b}, gets improved. However,
this kind of position is not equivalent with a player's individual power, but comes from the structure of
the network. In other models which could be used to simulate conflicts \cite{ra,szsz,adw,hk,eps}, it is only the
amount of actors of given orientation what matters for the final outcome of a model dynamics. Perhaps this
limitation is a legacy of statistical physics, where phase transitions are considered of a system of identical
objects. An extensive review of social simulations inspired by physics can be found in \cite{cfl}. \\

Despite its obvious validity for conflicts, dynamics of power of individual players has been ignored
in most computational models. As an exception, we note the Bonabeau model \cite{bn1} (note however,
that the term "Matthew effect" has not been used there). In this model, when two players meet they fight;
the winner gets more power and the loser - gets less. These gains and loses are relevant for the
outcome of subsequent fights. Main result of the Bonabeau model is a transition between egalitarian
and hierarchical phase of a model society, identified by means of simulations  and mean field modeling
\cite{bn2,bn3,bn4}.\\

Our aim here is to trace consequences of an individual strategy for the player who selected it. Hence the model
dynamics includes individual characteristics of players. Each player is endowed with the willingness to commit himself
to conflicts, which stands for his strategy, and with an initial value of the power. The former remains constant in time,
while the latter is a subject of model dynamics. Taking into account the principle 'rich gets richer' we can expect a clear
difference between winners and losers. As will be demonstrated below, the model outcome is that winner takes all.
The problem to solve is, how the distribution of the model parameters allows to appoint the winner.
Apart from the random assignment of the parameters among actors, the model is purely deterministic.\\

In the next section, the model is explained in details. Further, analytical results are presented in the form of
mathematical theorems and stability analysis of the model equations. These results are backed with
numerical calculations, shown in a separate section. Last section is devoted to the interpretation
and discussion.

%% ===========================================================================
\section{The model}
%% ===========================================================================

Let us denote the number of players by $m$, and the player index by $i=1,2,..,m$. The power of $i$-th player at time $t$ is $p_i^t$, and it is kept nonnegative.
The willingness of $i$-th player to commit into conflict is denoted by $c_i$, kept in the range $[0,1]$. The equation of motion is (cf. with \cite{KoTC,KoTC1})

\begin{equation}\label{de}
p_i^{t+1}=\frac{ p_i^t(1- c_i  r_i^t)}{z^t}, \  \  t=0,1,2,....
\end{equation}
where

\begin{equation}\label{der}
r_i^t =\frac{\sum_{k\neq i} p_k^t}{m-1} \equiv  \frac{1-p_i^t}{m-1}
\end{equation}
represents a mean player other than $i$-th one. The coupling between players is introduced {\it via} the normalization constant $1/z^t$, which is taken as to assure that

\begin{equation}\label{des}
\sum_{i=1}^m p_i^{t+1}=1,
\end{equation}
what marks that the total amount of power remains constant. In other words, we have a zero-sum game.\\

 Note that in general the law of conflict redistribution of power  is unknown. Our  rather simple version of conflicting fight presented by (\ref{de})  expresses the natural  primitive principle: {\it each against all}. Due to (\ref{des}) we find, that
\begin{equation}\label{den}  z^t=1-\theta^t, \ \ \theta^t:= \sum_{i=1}^m p_i^t c_i r^t_i.
\end{equation}
We will refer on (\ref{de}) as  the formula of  conflict interaction.

As we see, $p_i^{t+1}$ increases with $p_i^t$, what activates the Matthew effect. On the other hand, the whole contribution to $p_i^{t+1}$ from the conflict is negative. More precisely, $p_i^{t+1}$ grows, for players with $c_ir^t_i < \theta^t $, and falls, if $c_ir^t_i > \theta^t $.
Our computational problem is twofold:\\
\noindent
- what is the strategy $c_i$ which drives a player to success?\\
- how this strategy depends on the actual distribution of power?\\

To answer, we need a more deep mathematical analysis.

%% ===========================================================================

%% ===========================================================================
\section{Analytical results}

\subsection{Short analysis of the conflict formula }

  At first let us put $c_i=1$ for all $i=1,...,m$. Then using (\ref{der}) we can rewrite  (\ref{de}) in terms of coordinates of vector $p^t=(p_1^t,...,p_m^t)$ from $(m-1)$-dimensional simplex ${\bf S}^{m-1}_+$ as follows
\begin{equation}\label{2}
 p_i^{t+1} = p_i^t \cdot \frac{m-2 + p_i^t}{m-2+ L^t}, \ \
\end{equation}
where $L^t$ denotes the square norm of $p^t$, i.e.,
 \begin{equation}\label{Lt}
L^t\equiv \| p^t \|^2:=\sum_{k=1}^m ( p_k^t)^2.
\end{equation}
Thus
\begin{equation}\label{ka}
 p_i^{t+1} = p_i^t \cdot k_{i}^t, \ \ \ \  k_{i}^t:=\frac{m-2 + p_i^t}{m-2+ L^t}.
 \end{equation}
Now we observe that  if $p_i^t > L^t$,  then $k_{i}^t>1$ and therefore $p_{i}^{t+1}$ increases. $p_{i}^{t+1}$ will decrease, if    $p_i^t < L^t$.

Thus, the value  $L^t$  may be considered as  a threshold which divides the conflicting society  into three classes of players:
\begin{equation} \label{dec}
I^t_-:=\{i \ : p_i^t< L^t \}, \ \ I^t_0:=\{i \ : p_i^t= L^t \}, \ \ I^t_+:=\{i \ : p_i^t> L^t \}.
\end{equation}

It is easy to check that in general both subsets $I_-^t$ and \ $I_+^t, \  t\geq 0$ are always non-empty. In the excluding  case, when $  p_i^{t=0}=1/m, i=1,...,m   $,  \ $L^t=1/m$  too, and the set
 $I_0^t=m$    for all  $t \geq \infty$. In all other cases $I_0^t$ is non-empty only temporarily.

\subsection{A single winner is generic}
 For further  manipulations in situation $c_i=1$ for all $i$ we rewrite  (\ref{de}) in a form
\begin{equation}\label{22}
 p_i^{t+1}  =p_i^t \left(1+\frac{p_i^t-L^t}{m-2+L^t} \right)=(1+\delta_i^t) \cdot p_i^t,
\end{equation}
where
\begin{equation}\label{delta}
  \delta_i^t:=\frac{p_i^t- L^t}{m-2+ L^t}.
\end{equation}

  Three next propositions follows directly from  (\ref{2}) -- (\ref{delta}).

   Let us fix some initial distribution of power between players, i.e., we fix  $p\equiv p^{t=0}\in {\bf S}^{m-1}_+$, a stochastic vector  from  the positive simplex.

{\bf Proposition 3.1 }\label{ord} {\t If some couple  of initial coordinates satisfies   $p^{t=0}_i=p_k^{t=0}, i\neq k$, then  $p_i^t=p_k^t$ for all $t=1,2,...$ Moreover, if $p_i^{t=0}<p_k^{t=0}$, then  $p_i^t<p_k^t$ for all  $t\geq 1$.}

Thus,
\begin{equation}\label{ordN}\ p_i^t \leq p_k^t \Longrightarrow  p_i^{t+1} \leq p_k^{t+1}, \ \ t=0,1,...
\end{equation}

It means that
the conflict interaction does not change the initial ordering of players on their power:
 \begin{equation}\label{ordforall} 0 \leq p_{i_1}^{t=0}\leq p_{i_2}^{t=0}\leq \cdots p_{i_m}^{t=0}\leq 1  \Longrightarrow   0 \leq p_{i_1}^t  \leq p_{i_2}^t \leq \cdots p_{i_m}^t \leq 1, \ \ t=1,2,...
\end{equation}

In fact a sign of the difference  $p_{i}^t  - L^t$ in (\ref{22}) defines whether  $p_{i}^t$ grows or falls on $t+1$-step.

{\bf Proposition 3.2}\label{dyn}  If $p_{i}^t< L^t$, then
\begin{equation}\label{incr}\ p_{i}^{t+1}<  p_{i}^{t},
\end{equation}
and if $p_{i}^t> L^t$, then
\begin{equation}\label{decr}\ p_{i}^{t+1}>  p_{i}^{t}.
\end{equation}

{\bf Proposition 3.3}\label{4} The sequence   $L^t$   converges to a bounded limit:
\begin{equation}\label{b}\  0 < \lim_{t \rightarrow \infty} L^t =b \leq 1 .
\end{equation}

{\it Proof}. Obviously
$0 <  L^t \leq 1$, since  $L^t=\|p^t\|^2$ and all vectors  $p^t$ are stochastic. We have to show that the sequence  $ L^t$ is monotonically growing,
 \begin{equation}\label{incrN} L^{t+1}-L^t>0,
t\geq0.
\end{equation}
With this aim we use the decomposition (\ref{dec}). If $i' \in I_-^t$, then
${p^t_{i^\prime}}- L^t <0,\  \  \delta_{i^\prime}^t  <0.$  Therefore due to (\ref{22}) the difference $p_{i^\prime}^{t+1}-p_{i^\prime}^{t}=\delta_i^t\cdot p_i^t$ is negative. Denote it by $-d_{i^\prime}^{t}$ with $d_{i^\prime}^t>0.$
In the case $i'' \in I_+^t$ the opposite inequality is fulfilled,
$\delta_{i^{\prime \prime}}^t > 0$. Then $ p_{i^{\prime \prime}}^{t+1}-p_{i^{\prime \prime}}^{t}=:d_{i^{\prime \prime}}^t > 0.$

Since both $p^t$ and  $p^{t+1}$ are stochastic there exist    $ 0<s<m$ such that
$$0=\sum_{k=1}^m p_k^{t+1}-\sum_{k=1}^m p_k^{t}=\sum_{i^\prime=1}^s (p_{i^\prime}^{s+1}- p_{i^\prime}^{s})+ \sum_{{i^{\prime \prime}}=1}^{m-s} (p_{i^{\prime \prime}}^{s+1}- p_{i^{\prime \prime}}^{s})=
 - \sum_{i^\prime=1}^s d_{i^\prime}^t+\sum_{{i^{\prime \prime}}=1}^{m-s} d_{i^{\prime\prime}}^t.$$
By this
\begin{equation}\label{zero}\sum_{{i^{\prime \prime}}=1}^{m-s}d_{i^{\prime\prime}}^t-\sum_{i^\prime=1}^s d_{i^\prime}^t=0.
\end{equation}
Consider now the difference $L^{t+1}-L^t \equiv \|p^{t+1}\|^2 - \|p^t\|^2$. Since $p_{i^\prime}^{t+1}=p_{i^\prime}^{t}-d _{i^\prime}^{t}$ and $p_{i^{\prime \prime}}^{t+1}=p_{i^{\prime \prime}}^{t}+d_{i^{\prime\prime}}^t$,  using the inequalities    $p^t_{i^{\prime\prime}} > \| p^t \|^2$ \ ($i^{\prime\prime} \in I^t_+$) and $p_{i^{\prime }}^{t} <  \|p^t\|^2$ ($i^{\prime} \in I^t_-$) we obtain
$$\|p^{t+1}\|^2 - \|p^t\|^2 > 2 L^t (\sum_{{i^{\prime \prime}}=1}^{m-s}d_{i^{\prime\prime}}^t-\sum_{i^\prime=1}^s d_{i^\prime}^t)+\sum_{i^\prime=1}^s {d_{i^\prime}^t}^2+\sum_{{i^{\prime \prime}}=1}^{m-s}{d_{i^{\prime\prime}}^t}^2>0.$$
Finally due to (\ref{zero}) we have:
$$L^{t+1}-L^t=\sum_{i^\prime=1}^s {d_{i^\prime}^t}^2+\sum_{{i^{\prime \prime}}=1}^{m-s}{d_{i^{\prime\prime}}^t}^2>0.$$
Thus
(\ref{incrN}) is proved. This shows that $L^t$ is a growing bounded sequence. Therefore   (\ref{b}) is true.

Let us denote
$$p_{\rm max}^t:=\max_{1\leq i \leq m } \{ p_i^t\}.$$
Now we will prove one of the main result of the paper.

{\bf Theorem 3.1} Assume for a  vector $p\equiv p^{t=0} \in {\bf S}^{m-1}_+, \ m>2$ all its coordinates are non-zero and mutually different,
\begin{equation}\label{inj} p_i \neq p_j, \ \ i \neq j.
\end{equation}
Then
\begin{equation}\label{limall} \lim_{t\to \infty} p_i^t=0, \ \ p_i\neq p_{\rm max}.
\end{equation}
and
\begin{equation}\label{limmax} p_{\rm max}^\infty:=\lim_{t\to \infty} p_{\rm max}^t=1,
\end{equation}

{\it Proof}.  From obvious inequalities
\begin{equation}\label{neq2} \min_k \{p_k^t\} \leq \|p^t\|^2 \equiv L^t \leq  \max_k \{p_k^t \}
\end{equation}
and  Propositions 3.2, 3.3 it follows that the sequence   $p_{\rm max}^{t}$ grows with  $ t\rightarrow\infty$. Since it is bounded, there exists a limit   $a=\lim_{N\to \infty} p_{\rm max}^{t}  \leq 1$. Due to condition (\ref{inj}) without loss of generality we can assert that  coordinates of vectors  $p^t$ are ordered in such a manner that
\begin{equation}\label{order}
0<p_1^t <p_2^t < \cdots <p_m^t<1.
\end{equation}
This order does not depend from  $t$ (see Proposition 3.1). By this the latter coordinate is maximal for all times   $p_{\rm max}^{t}=p_m^t$. Thus, the following estimate holds: $$0< a:=\lim_{t\to \infty}p_m^t= p_m^\infty\leq 1.$$
Let us prove that $  a = 1.$ At first we show that $a=b$, where $b=L^\infty:=\lim_{t \rightarrow \infty} L^t$. Indeed,  from existence of the limits for  $p_m^t$ and $L^t$ we have:
\begin{equation}\label{a}
a=p_m^\infty=k^\infty_m \cdot p_m^\infty=\frac{m-2+p_m^\infty}{m-2+L^\infty}\cdot p_m^\infty=\frac{m-2+a}{m-2+b}\cdot p_m^\infty.\end{equation}
By this $k^\infty_m=1.$
It means that $a=b$. In fact both, $a$ and $b$ are equal to one. This consequence one can draw from analysis of behavior of the lasting
 coordinate $p^t_{m-1}$. Indeed, due to (\ref{order}) and Proposition 3.1, the inequality $p_{m-1}^t< p_{m}^t$  holds for all $t$. Therefore the ratio $p_{m}^t/ p_{m-1}^t$ exceeds one and grows.  In particular,   $p_{m-1}^t< a$ always. It means that this ratio goes to infinity and therefore  $\lim p_{m-1}^t=0$.  If we assume the contrary, then by (\ref{ka}) we have the equality $\lim p_{m-1}^t=L^\infty=b=a$, that is a contradiction.
 Similarly one can assert that all other coordinates  converge to zero,  $\lim p_{i}^t=0, \ i\neq m$. Thus,  $a=1$. The theorem is proved.

 The Figure 1  illustrates the above result.

\begin{figure}[h] \centering
\includegraphics[scale=0.2]{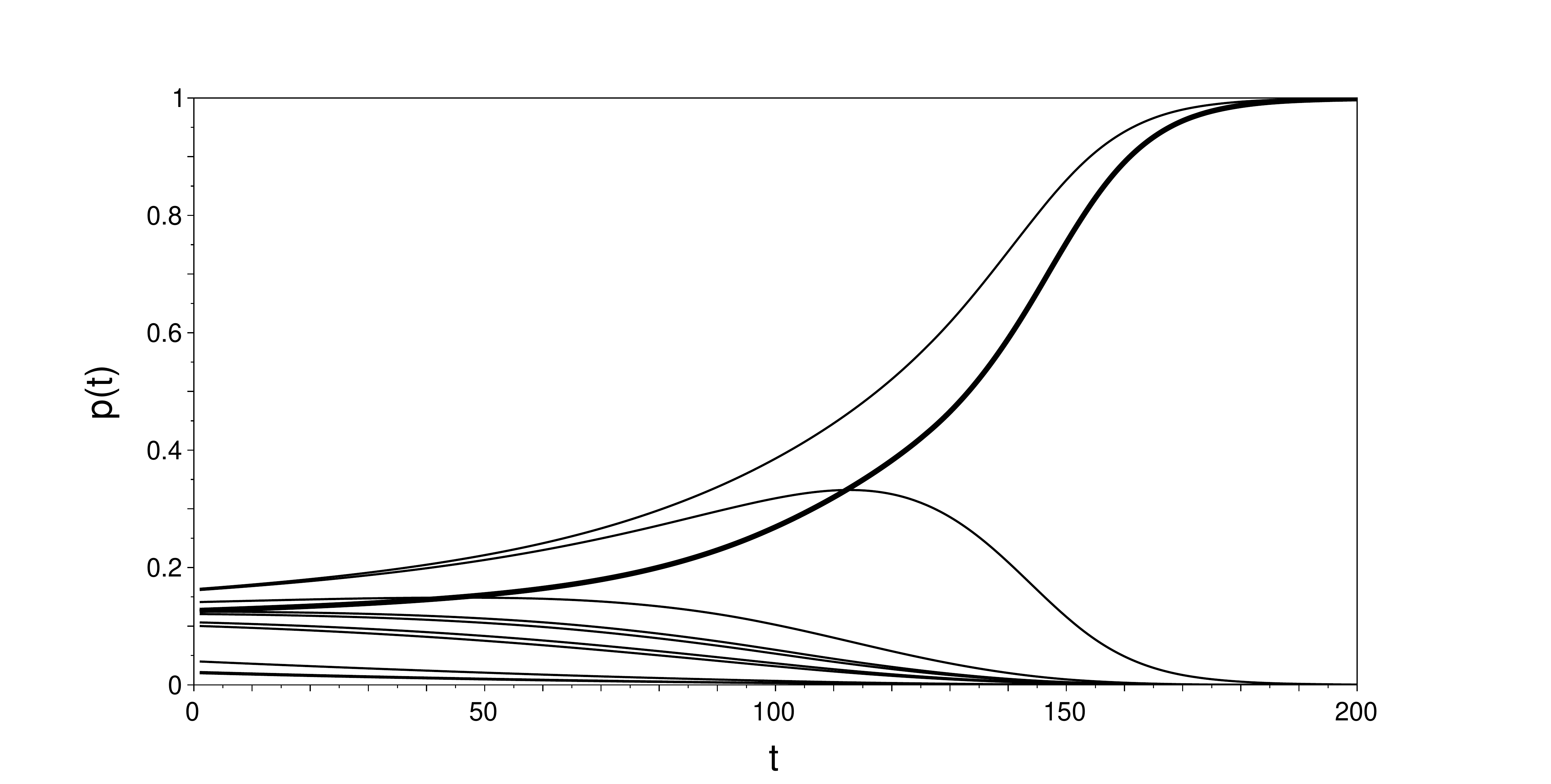}
\caption{{The winner is determined by the maximal initial value of social
power.  $m=10, c_i =1 $ for all $i= \overline{1,10}$.  The initial values are
: $p_1=0.05, p_2=0.01, p_3=0.01, p_4=0.02, p_5=0.053, p_6=0.08, p_7=0.06,
p_8=0.07, p_9=0.062, p_{10}=0.011. $ Bold line shows  the growth of $L^t$. }}
%\label{Figure-KK1}
\end{figure}

By this theorem if all initial coordinates $p_i^{t=0}$  are different, then
 $$| I^t_- | \longrightarrow m-1, \ \  | I^t_+ | \longrightarrow 1, \ \ t \longrightarrow \infty,$$
 where $| I^t_{\rm sign}|$ denotes a cardinality of a set.

Thus  we proved that if the society  joints players with nonzero and different values of power, there exist   a single winner  which  is determined  by the maximal initial power. In other terms this means  that the richest  player  becomes  richer and captures with time the whole wealth, all other agents do not get anything.\\

Let ${\cal P}^*$ denotes the set of fixed points for the nonlinear map
generated by formula (\ref{de}). By construction, all limit points in Theorem
3.1 are fixed points, write $ p^\infty = p^* \in  {\cal P}^*_1$, where index
$1$ means that  $ p^* $ has only one nonzero coordinate equals to  $1$.\\

{\bf Theorem 3.2} Under condition (\ref{inj}) there exist $m$ fixed points \begin{equation}\label{fpm} p^*_j=(0,0,...,0,p_{ji},0,...,0), \ \ p_{ji}=\delta_{ji}, \ \
j,i=1,...,m,
 \end{equation}
 where $\delta_{ji}$ stands for the Kronecker symbol. All these points are stable.

{\it Proof}. We have only to show the stability of the fixed points $
p^\infty_j = p^*_j, j=1,...,m$. It follows from the fact that any
$\varepsilon$-perturbation of the vector
$p^*_j=(0,0,...,0,\delta_{ji},0,...,0)$ preserves for its $j$-coordinate to
have the maximal value. And by Theorem 3.1 the limits on $t$ for all other
coordinates are zero.\\

Consider the exotic situation when initial vector  $p\in {\bf S}^{m-1}_+$ has
$1<k<m$ equal coordinates with the maximal value. Obviously the set of such
vectors has zero $(m-1)$-dimensional Lebesgue measure. By slightly modified
argumentations as above one can prove that all non-maximal coordinates
$p_{i\neq \rm max}^t$ converge to zero, as $t\to \infty$, and coordinates with
maximal value come to $1/k$.
 Thus, the limiting set of fixed points, denote it by ${\cal P}^*_k$, contains the family of  $C^m_k$  vectors   $\{ p^* \equiv p^\infty \}$ whose $k$ nonzero coordinates are equal to
 $1/k$.\\

{\bf Theorem 3.3} Every fixed point from family   ${\cal P}^*_k, 1< k \leq m$
is unstable.

{\it Proof}. Obviously, a general $\varepsilon$-perturbation of a vector
$p^*\in {\cal P}^*_k$ does not preserve the condition that $k\geq 2$
coordinates are equal and  have the maximal value. Therefore by Theorem 3.1 the
limiting vector will not belong to  $p^*\in {\cal P}^*_k$.

\subsection{An arbitrary conflict activity}

 Consider the general situation when $0\leq c_i \leq 1$ are arbitrary.

In this case the conflict  formula  (\ref{de}) after  using (\ref{der})  has a view
 \begin{equation}\label{2c}
 p_i^{t+1} =  p_i^t \cdot \frac{m-1 -c_i (1-p_i^t)}{m-1 - L^t_c}= p_i^t \cdot k_{i,c}^t \ \
\end{equation}
where
\begin{equation}\label{kac}
k_{i,c}^t:=\frac{m-1 -c_i (1- p_i^t)}{m-1- L^t_c}
 \end{equation}
 and
 \begin{equation}\label{Ltc}
L^t_c:=\sum_{k=1}^m c_k p_k^t (1-p_k^t).
\end{equation}
Obviously now  the value  $0 \leq L^t_c \leq 1 $  has more complex  non-linear dependence from $p_i^t$, in particular, it changes non-monotonically with time.

 In a slightly other form the conflict  formula  views as follows:
\begin{equation}\label{22c}
 p_i^{t+1}=p_i^t\left(1+\frac{L^t_c - c_i (1-p_i^t)}{m-1 -L^t_c}\right)= p_i^t (1+\delta_{i,c}^t),
\end{equation}
where
\begin{equation}\label{deltac}
  \delta_{i,c}^t:=\frac{L^t_c - c_i (1-p_i^t)}{m-1 -L^t_c}.
\end{equation}

From (\ref{2c}) and  (\ref{22c}) we see that $p_i^t$ increases under the following condition
\begin{equation}\label{condc}
L^t_c>  c_i(1-p_i^t).
\end{equation}
Let
\begin{equation}\label{Ltic}
L^t_{i,c}:=\sum_{k \neq i} c_k p_k^t (1-p_k^t),
\end{equation}
then (\ref{condc}) has a form
\begin{equation}\label{condc2}
   \frac{L^t_{i,c}}{ c_i}> (1-p_i^t)^2.
\end{equation}

Unfortunately, in general, no one of both conditions (\ref{condc}),  (\ref{condc2}) guarantee the global increasing for  $p_i^t$, but only the local behavior.
Nevertheless, we are able to get some strategic characteristic of the relative behaviours for players in terms of their ratios
$$R^t_{ik}:=\frac{p_i^t}{p_k^t}, \ \ i,k=1,...,m.$$
Since due to (\ref{2c})
  \begin{equation}\label{Rik}
  R^{t+1}_{ik}=R^t_{ik}\cdot \frac{m-1-c_i(1-p_i^t)}{m-1-c_k(1-p_k^t)},
   \end{equation}
 we obtain

 {\bf Proposition 3.4} The ratio $R^t_{ik}$ grows with $t\to \infty,$ iff
 \begin{equation}\label{cpik}
 c_i (1-p_i^t) < c_k (1-p_k^t).
 \end{equation}

{\bf Theorem 3.4}  Assume
\begin{equation}\label{i0Lck}
 c_i (1-p_i^t) < L_c^t < c_k (1-p_k^t).
 \end{equation}
 hold for a single $i=i_1$ and all $k\neq i_1$.
Then
\begin{equation}\label{i11}
p_{i_1}^\infty = \lim_{t \to \infty} p^t_{i_1} = 1, \ \ p_{k \neq i_1}^\infty = \lim_{t\to \infty} p^t_{k} =0. \end{equation}
All these
limit points are stable.

{\it Proof}. By the left part of  (\ref{i0Lck}),  $p_{i_1}^{t+1}$ grows  (see (\ref{22c}).  Obviously $R^{t+1}_{i_1k}> 1$ since due to (\ref {i0Lck}), we have
 \begin{equation}\label{cpik}
 c_{i_1} (1-p_{i_1}^t) < c_k (1-p_k^t).
\end{equation}
 and therefore the inequalities
\begin{equation}\label{cpikt+1}
c_{i_1} (1-p_{i_1}^{t+1}) < c_k (1-p_k^{t+1})
\end{equation}
are also true. They, in general, do not guarantee that $p_{i_1}^{t+1}$ grows
quicker than each   $p_{k}^{t+1}$. But thanks to   the right part of (\ref {i0Lck}),  all    $p_{k}^{t+1}$ in fact decrease. Since (\ref {i0Lck}) are fulfilled for each
$t$ we get ({\ref{i11}}). Clearly, the limit points are fixed.

To prove its stability we will consider without of loss generality the case $p^\infty \equiv p^*=(1,0,0,...,0)$ and show this vector attracts all vectors of type $p^{*,\varepsilon}=(1-\varepsilon_1,
\varepsilon_2,...,\varepsilon_m), \ \varepsilon_1=\sum_{k\neq 1} \varepsilon_k$ with $\varepsilon_1$ small enough. In fact, we have to check the inequality $(p_1^{*,\varepsilon})^{t=1} > p_1^{*,\varepsilon}= 1-\varepsilon_1$ only for  the first coordinate. It is  equivalent (see  (\ref{i0Lck})) to $$ c_1(1-p_1^{*,\varepsilon}) < L_{c,\varepsilon}, \
L_{c,\varepsilon}=\sum_k c_k \varepsilon_k(1-\varepsilon_k).$$
In turn, the equivalent inequality has a form $c_1\varepsilon_1^2<\sum_{k\neq 1} c_k  \varepsilon_k(1-\varepsilon_k)$, or
$\sum_k c_k  \varepsilon_k^2 < \sum_{k\neq 1} c_k \varepsilon_k $. Clearly, the last inequality is fulfilled for all $\varepsilon_k$ small enough, since the left side is constituted with square of  $\varepsilon_k$.\\

{\bf Proposition 3.5} Assume $c_{i_1}<c_k$ and for some $t$ the inequalities
 \begin{equation}\label{i0Lckk}
 c_i (1-p_i^t) < L_c^t < c_k (1-p_k^t), \ \  p^t_k<1/2
 \end{equation}
 hold for a single $i=i_1$ and all $k\neq i_1$. Then these inequalities are true for all $t+n, n\geq 1$.

{\it Proof}.  Obviously $R^{t+1}_{i_1k}> 1$ since due to (\ref {i0Lckk}) we have  (\ref{cpik}) for     $i=i_1$ and all $k\neq i_1$ Therefore the inequalities (\ref{cpikt+1}) hold too.
To show $c_{i_1}(1- p_{i_1}^{t+1}) < L_c^{t+1}$ one can consider the ratio
$$\frac{L_c^{t+1}}{c_{i_1}(1-p_{i_1}^{t+1})}=\frac{c_{i_1}p_{i_1}^{t+1}(1-p_{i_1}^{t+1})}{c_{i_1}(1-p_{i_1}^{t+1})}+ \frac{\sum_{k\neq i_1} c_{k}p_{k}^{t+1}(1-p_{k}^{t+1})}{c_{i_1}(1-p_{i_1}^{t+1})}.$$
Using (\ref{cpikt+1}) we find that $L_c^{t+1}/c_{i_1}(1-p_{i_1}^{t+1})> \sum_{i=1} ^m p_i^{t+1}=1 .$ Therefore $p_{i_1}^{t+1}$ grows.
The proof of  validity $L_c^{t+1} < c_k (1-p_k^{t+1})$ requires more deep observations. At $(t+1)$-th step the value of $L_c^{t}$ changes due to two reasons. At first, it falls since all $p^{t+1}_k$  falls due to   assumption  $p^t_k<1/2$ and by  inequalities  $L_c^t < c_k (1-p_k^t)$ (see (\ref{22c}) with $i=k$). At second, it grows since $c_{i_1} (1-p_{i_1}^t) < L_c^t$ (see also (\ref{22c}) with $i=i_1$). We assert that  inequalities $L_c^{t+1} > c_k (1-p_k^{t+1})$ could not fulfilled if $c_{i_1}< c_k$ for all $k\neq i_1.$ The proof is purely geometrical.  To show this fact one need to compare the graphics of functions   $c_{i_1} p_{i_1}^t (1-p_{i_1}^t) $ and
$c_{k} p_{k}^t (1-p_{k}^t) $ for $k$ with maximal value of $p_{k}^t$. By induction we continue our argumentations  for any $n>1$.

Thus,  (\ref{i11})  is also true under conditions of Proposition 3.5 (see Figure 2).\\

\begin{figure}[h] \centering \includegraphics[scale=0.2]{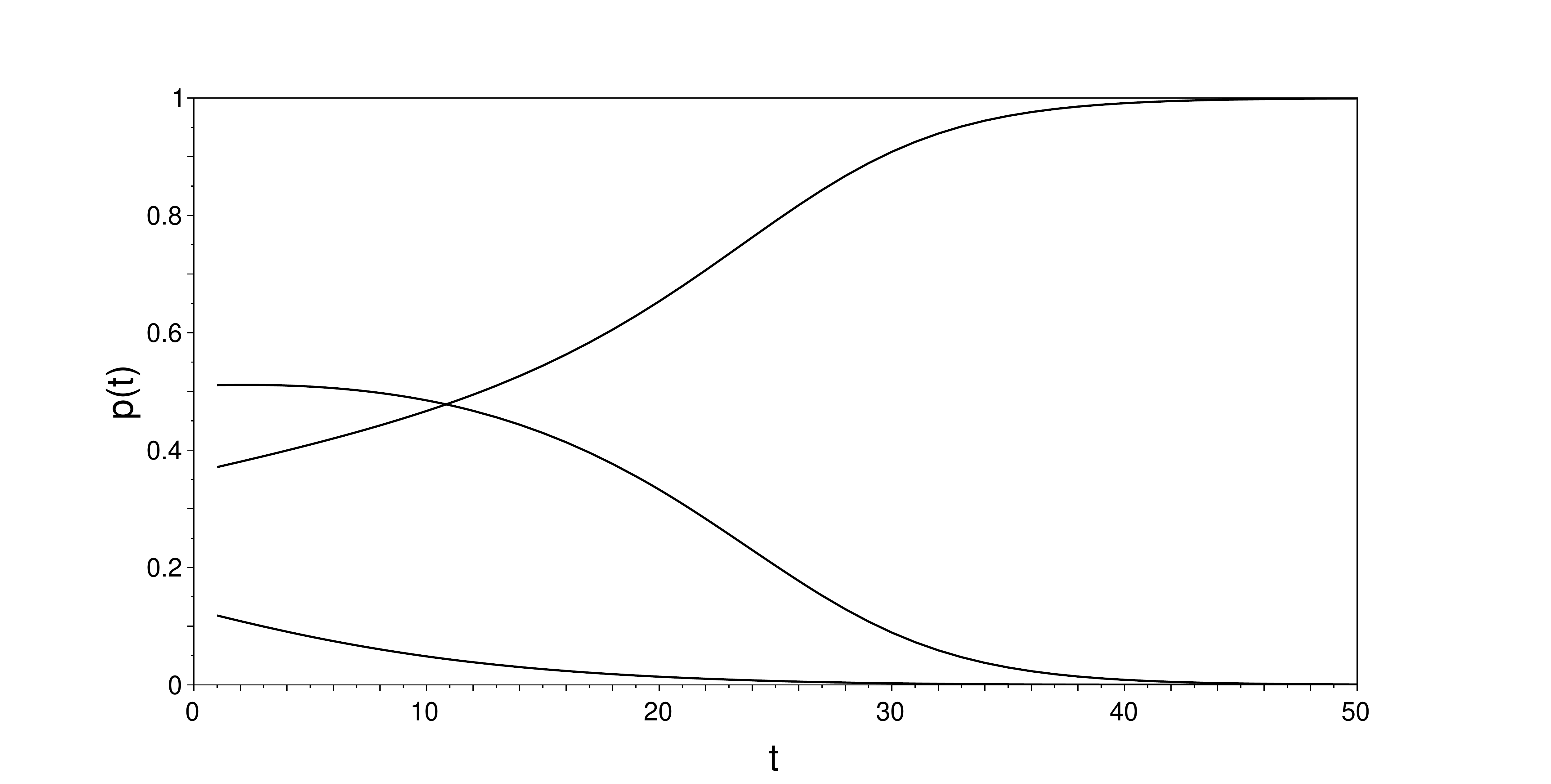} \caption{{  The second
player becomes winner due to conditions (\ref{i0Lckk}) from Proposition 3.5. $m=3, p_1=0.4065, p_2=0.2513, p_3=0.3421$, $c_1=0.4588, c_2=0.41967, c_3=0.2896 $ }}
\end{figure}

  Denote  by ${\cal P}^*_{k,c}$ the set of  fixed points for the general case $0 \leq c_i \leq 1$, where $1 \leq k \leq m$ means a number of nonzero coordinates.\\

 {\bf Theorem 3.5} Each fixed point $p^* \in {\cal P}^*_{k,c},  k >1$ is unstable.

 {\it Proof}  Consider any   $p^* \in {\cal P}^*_{k,c},  k>2$ and a couple of it nonzero coordinates  $p^*_{i_1}, p^*_{i_2} $. They have to satisfy the equality
 $$c_{i_1}(1-p^*_{i_1})= L_{c}^*=c_{i_2}(1-p^*_{i_2}).$$
Assume $p_{i_1}^* \geq p_{i_2}^*$ and replace  $p_{i_1}^*$ on  $p_{i_1, \varepsilon}^* =p_{i_1}^* + \varepsilon $ and  $p_{i_2}^*$ on  $p_{i_2, \varepsilon}^* =p_{i_1}^*- \varepsilon, \varepsilon >0 $.
 Then we easily  check that for any small $\varepsilon$, the above equalities transform into inequalities
$$c_{i_1}(1-p^*_{i_1, \varepsilon} )< L_{c, \varepsilon}^* < L_c< c_{i_2}(1-p^*_{i_2, \varepsilon}).$$
Since now $p^*_{i_2, \varepsilon}<1/2$  due to $k>2$, we can use Proposition 3.5. Thus,  $(p^*_{i_1, \varepsilon})^t$ increases and $(p^*_{i_2, \varepsilon})^t$ falls. For the case $k=2$ see next subsection.

\subsection{Structure of fixed points}

For clarity our assertion in more details, first  we consider the case $m=2$; although very simple, it provides a good starting point. The set of Eqns (\ref{de}) reduce to only one

\begin{equation}\label{m2evo}
p^{t+1}=\frac{p^t(1-c_1+c_1 p^t)}{1-(c_1+c_2) p^t (1-p^t)}.
\end{equation}
with three fixed points: $p^*=0$, $p^*=1$ and $p^*=c_1/(c_1+c_2)$. To state their stability, we need to compare $|\partial p^{t+1}/\partial p^t|$, calculated
at the fixed point, with one \cite{glen}. This expression is equal to $1-c_1$ and $1-c_2$ for $p^*=0$ and $p^*=1$, respectively. Hence, both these fixed points are stable except
the cases $c_i=0$, where the stability is marginal. At the third fixed point the derivative is $|\partial p^{t+1}/\partial p^t|=(c_1+c_2)/(c_1+c_2-c_1c_2)>1$, hence this fixed point is unstable. This is an illustration of the above-given general theorem, that all fixed points different than $\{p_i^*\}=\{0,0,...,0,1,0,...,0\}$ are unstable.\\

It is easy to see that the attraction basin for the fixed point $p*=1$ \ (p*=0) is interval $(c_1/(c_1+c_2), 1]$ \ ($[0,c_1/(c_1+c_2) )$. Indeed, since now $L_c=(c_1+c_2)p(1-p)$ from (\ref{22c}) it follows that $p^t$ grows to $1$ only if $c_1(1-p)< L_c$, i.e., if $p>c_1/(c_1+c_2)$. Otherwise, i.e., if  $p<c_1/(c_1+c_2)$, that is equivalent to
$c_1(1-p)> L_c$, the value $p^t$ goes to zero.  The unstable fixed point $c_1/(c_1+c_2)$ has its basin empty.

One can put the inverse question (some kind of the two players problem). Given
$0<p<1$ and $0<c_2\leq 1$ what  $c_1$ guarantees $p*=1$? From  (\ref{22c}) we
find solution $c_1< c_2 p/(1-p)$. In particular, if $c_2=1$ the first player
with any initial $p>0$  wins if he take   $c_1<  p/(1-p)$.

We note, in the case of three players the similar question (see below) requires essentially more effort.

For $m=3$, the normalization condition reduces the number of equations to two. For simplicity, let us use variables $x,y,1-x-y$ instead of $p_1,p_2,p_3$, and primes instead of time
index $t+1$; the time index $t$ will be omitted. Then we have

\begin{eqnarray}\label{m3evo}
x'&=&\frac{x(1-c_1+c_1 x)}{1-c_1 x(1-x)-c_2y(1-y)-c_3(x+y)(1-x-y)}\\
y'&=&\frac{y(1-c_2+c_2 y)}{1-c_1 x(1-x)-c_2y(1-y)-c_3(x+y)(1-x-y)}\nonumber
\end{eqnarray}
Basically, there are seven fixed points: $(x^*,y^*)$ = {\it i)} $(1,0)$, {\it ii)} $(0,1)$, {\it iii)} $(0,0)$, \\
{\it iv)} $(c_1/(c_1+c_2),c_2/(c_1+c_2))$, {\it v)} $(c_1/(c_1+c_3),0)$, {\it vii)} $(0,c_2/(c_2+c_3)$, and {\it vii)}

\begin{equation}\label{m3fp7}
\Big(\frac{c_1(c_2+c_3)-c_2c_3}{c_1c_2+c_2c_3+c_3c_1},\frac{c_2(c_3+c_1)-c_1c_3}{c_1c_2+c_2c_3+c_3c_1}\Big)
\end{equation}
However, the coordinates of the last fixed point are not necessarily positive. To keep all coordinates $(x^*,y^*,1-x^*-y^*)$ nonnegative, three conditions should be fulfilled:

\begin{eqnarray}\label{m3int}
c_1 & > & c_2c_3/(c_2+c_3)\nonumber \\
c_2 & > & c_3c_1/(c_3+c_1) \\
c_3 & > & c_1c_2/(c_1+c_2) \nonumber
\end{eqnarray}
Now suppose that with the coefficient $c_3$ we are at the limit case, {\it i.e.} $c_3(c_1+c_2)=c_1c_2$. After some simple algebra we get $x+y=1$, hence for the seventh fixed point given by Eq. (\ref{m3fp7}) we get $p_3^*=0$. Also, its first coordinate $x^*=c_1/(c_1+c_2)$, what means that the two fixed points $\it (iv)$ and $\it (vii)$ collide. When $c_3$ decreases further, the seventh fixed point leaves the simplex where $\{p_i>0\}$.\\

As we know from the preceding subsection, the only stable fixed points appear at the corners of the $m$-cube, where one player got the whole power ($p_i^*=1$). We can check the stability of such fixed points, taking $x=1$ as an example. There, the eigenvalues of the Jacobian are ($1-c_3/2$, $1-c_2/2$), what is nicely consistent with the case $m=2$.\\

The question about the attraction basins for stable fixed points is more complex and here we present only particular numerical results.

Consider the case {\it i}), two next cases,  {\it ii}), {\it iii}) are analogical.

Let $c_1<c_2, c_3$. If $p_1>p_2, p_3$, then,  due to (\ref{Rik}), the both inequalities  $c_1(1-p_1)< c_2(1-p_2), c_3(1-p_3)$ become with time stricter. This means that $p_2', p_3'$ fall    and hence  $x^*=1$. It is only a part of the attractive basin for point $x^*=1$. Conditions $p_1=p_2>p_3$,  $p_1=p_3>p_2$ give, by same argumentation, else two parts. Moreover, for  enough small $c_1$ the attractive basin of $x^*$ contains points $p_1, p_2, p_3$ with $p_1< p_2, p_3$. It follows from the fact that  if $c_1=0$, then  $x^*$ attracts all  points with $p_1\neq 0$ since in this case both ratios $R_{1,2}^t, R_{1,3}^t$ grows (see (\ref{Rik}). Figure 3 demonstrates above phenomenon for $p_1$ with minimal value in a general case (m=10).

\begin{figure}[h] \centering
\includegraphics[scale=0.2]{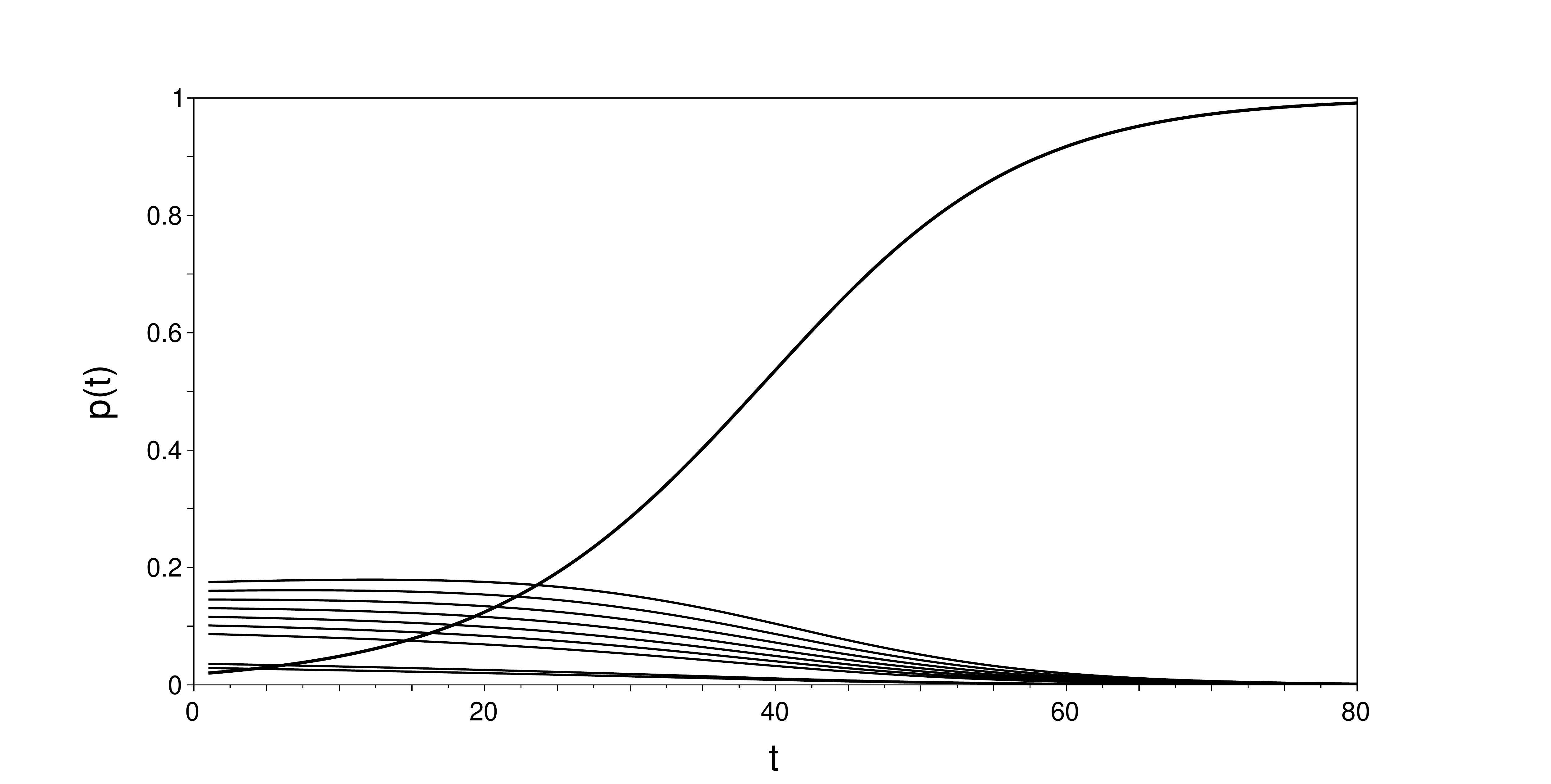}
\caption{{The player with smallest initial social energy (power) becomes the
winner. $m=10, c_k=0, p_k= \min_i \{p_i\}, c_i =1 $ for all $i=
\overline{1,10}, i\neq k.$
   The initial values are: $p_1=0.0182,
p_2=0.0364,
p_3=0.1309,
p_4=0.0290,
p_5=0.1164,
p_6=0.1018,
p_7=0.1455,
p_8=0.0873,
p_9=0.16,
p_10=0.1745.$
  }}
%\label{Figure-KK1}
\end{figure}

The fixed points {\it iv}), {\it v}), {\it vi}) are highly unstable. Any small perturbation of points x*, y* violates the balance $c_1(1-x*)/c_2(1-y*)=1$ which, due to (\ref{Rik}), goes far from $1$ with time.

In the general case of unlimited number $m$ of players and $c_i>0$ $\forall i$, we have $m$ 'corner' fixed points and $m(m-1)/2$ 'edge' fixed points where $p_i^*=0$ for all but two players. As we have seen for $m=3$, more fixed points are possible if the coefficients $c_i$ fulfill appropriate conditions. Accordingly, the maximal number of the fixed points is

\begin{equation}\label{maxfp}
\sum_{k=1}^m \binom{m}{k}=2^m-1.
\end{equation}
The actual number of fixed points can be less, if some of them fall out of the area where $\forall i, p_i > 0$.\\

The positions of the unstable fixed points can give hints on the boundaries of the basins of atraction of the stable fixed points. This advantage is demonstrated numerically in the next section.

%% ===========================================================================
\section{Basins of attraction: $m=3$ and beyond}
%% ===========================================================================

\begin{figure}[h] \centering
\includegraphics[scale=0.6]{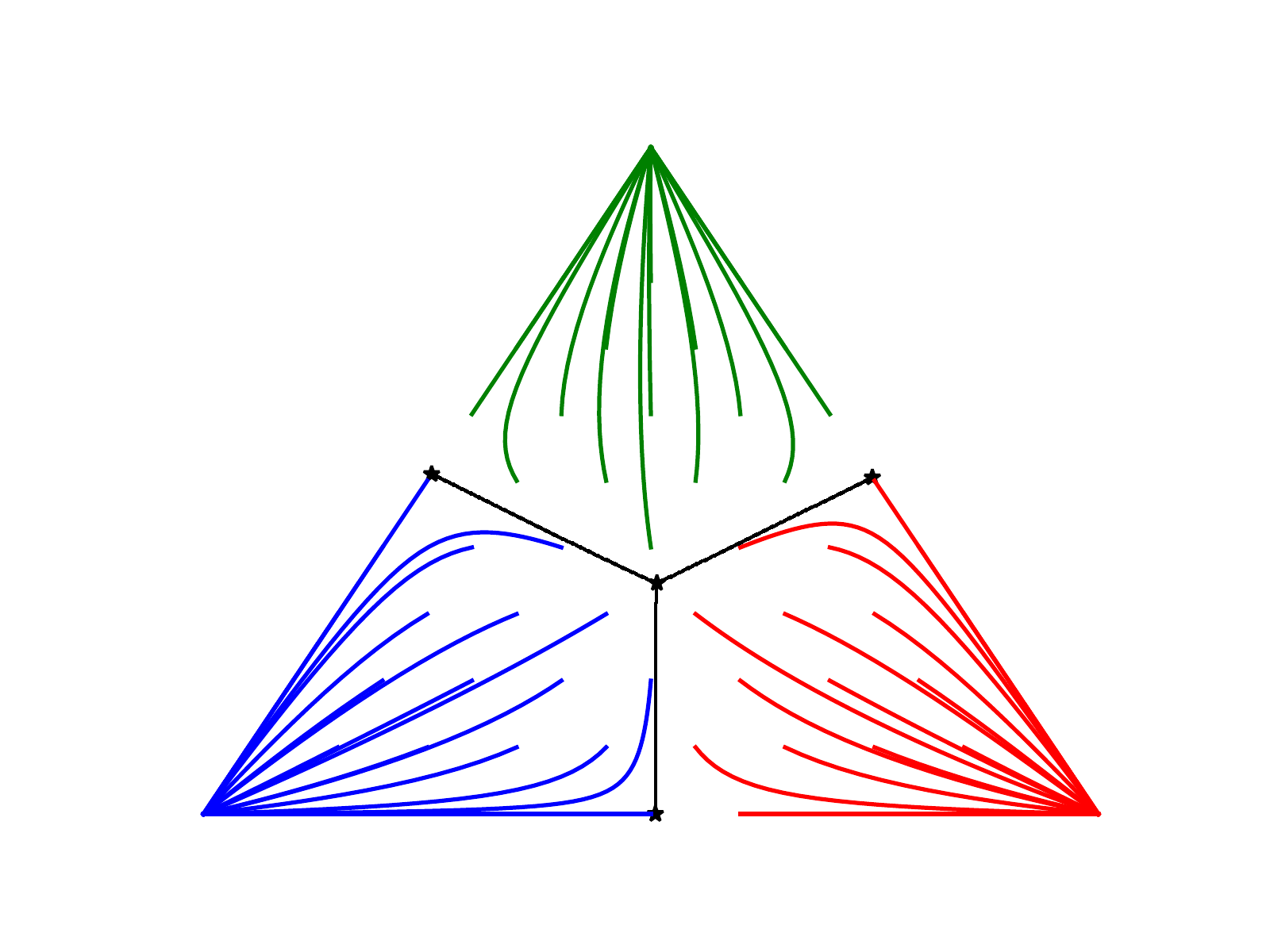}
\caption{{For close values of conflict activities,  $c_1=0.5,
c_2=0.49,
c_3=0.51,$  basins of attraction divide the $2$-dimensional simplex into three parts of similar size. The unstable fixed points, marked here by stars, lie at the boundaries between the basins, marked by continuous black lines. Red trajectories (color online) tend to $(1,0,0)$ (bottom right), blue trajectories tend to $(0,1,0)$ (bottom left), and green trajectories tend to $(0,0,1)$ (top). }}
\label{tt1}
\end{figure}

\begin{figure}[h] \centering
\includegraphics[scale=0.6]{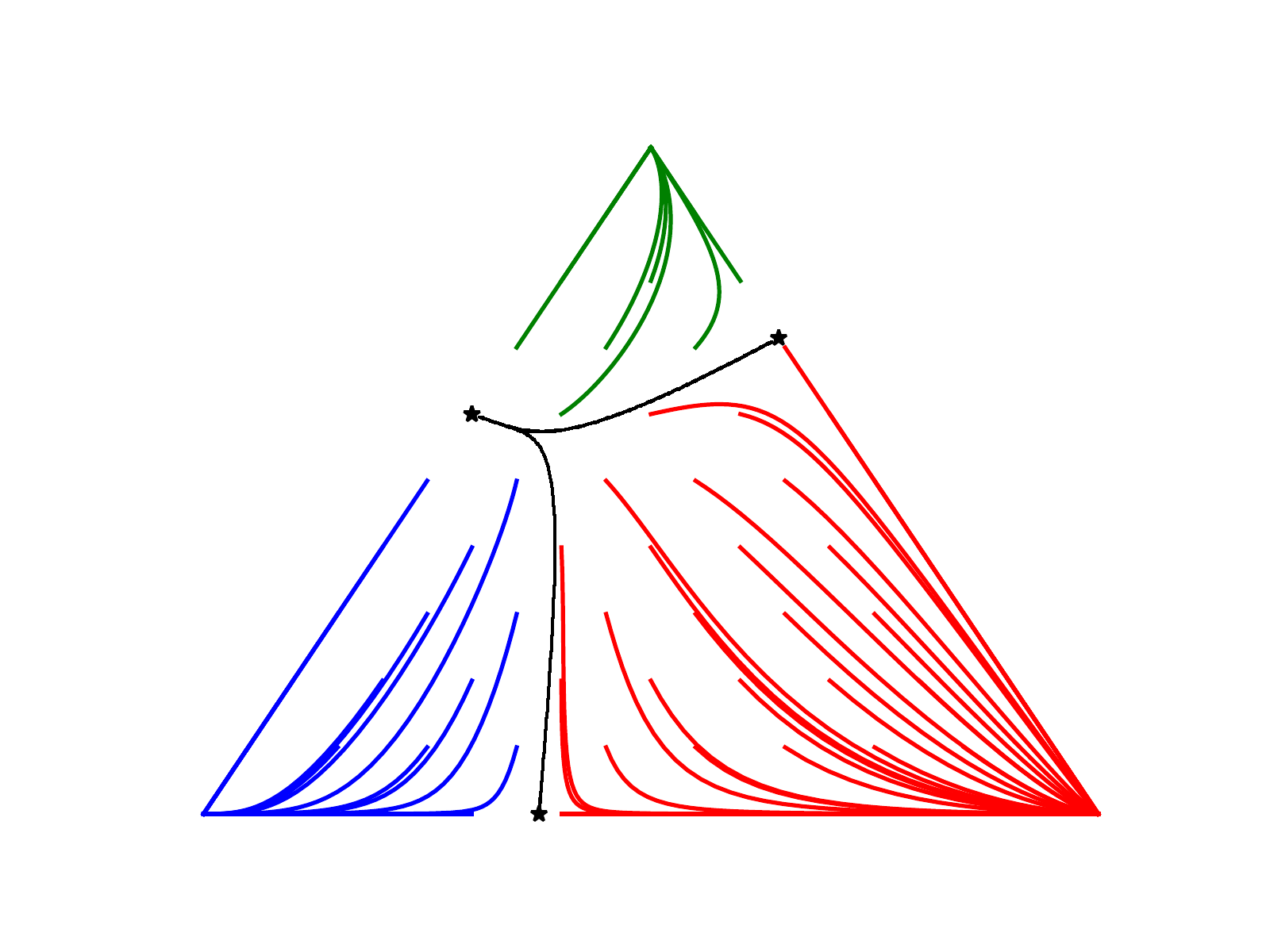}

\caption{{The fixed point $p^*=(1,0,0)$ (bottom right) which represents the player with the smallest conflict activity ($c_1=0.24,
c_2=0.4,
c_3=0.6,$)  has the largest basin of attraction. Here the central (seventh) fixed point collides with one of the edge fixed points.}} 
\label{tt2}
\end{figure}

\begin{figure}[h] \centering
\includegraphics[scale=0.6]{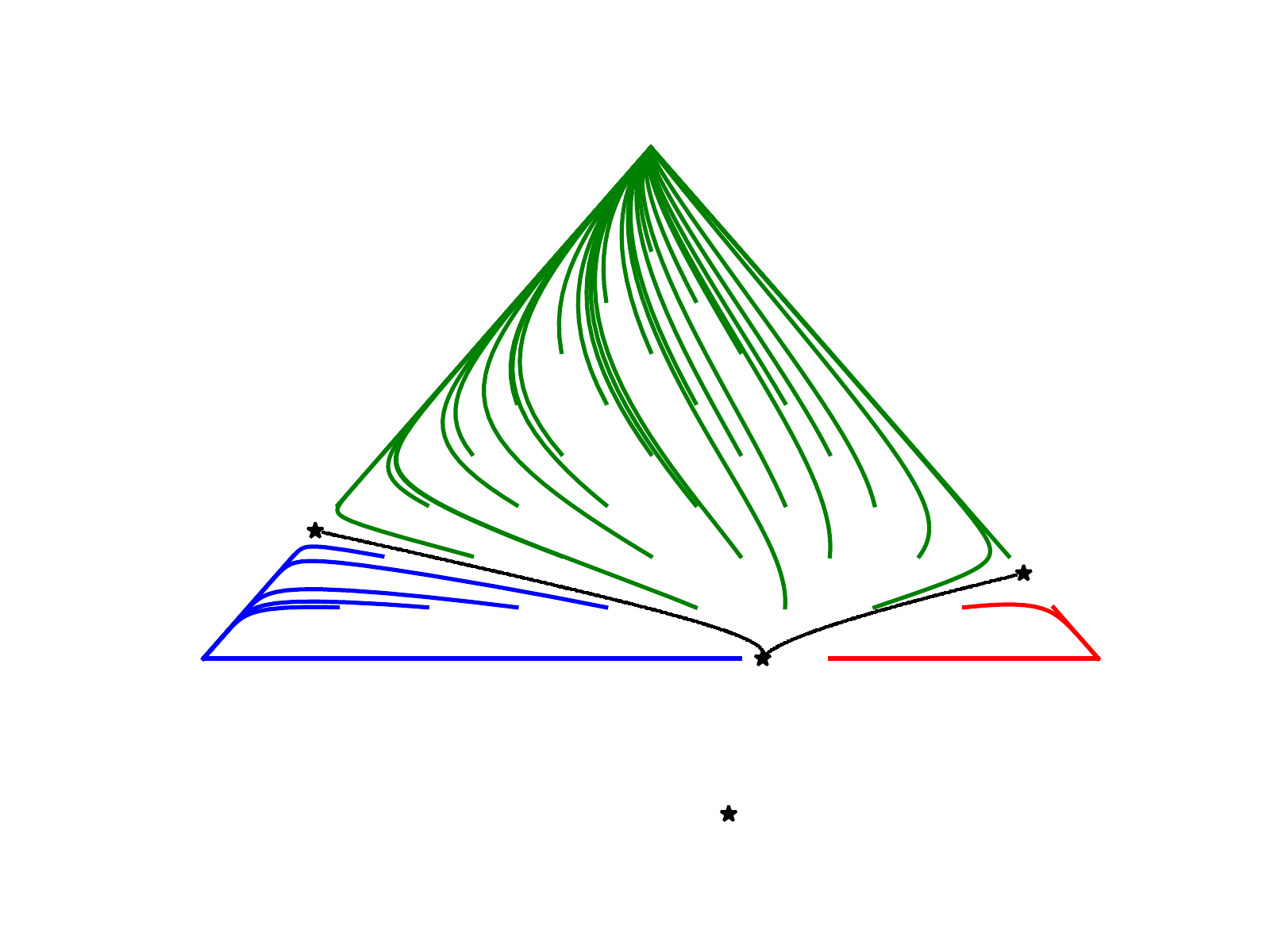}

\caption{{For activity values $c_1=0.5,
c_2=0.3,
c_3=0.1 , y=p_3$ the central (seventh)
fixed point is arranged  out of the simplex and the basins of attraction are subjected to a strong deformation.}}
  \label{tt3}
\end{figure}

In Figs. (\ref{tt1} , \ref{tt2} , \ref{tt3}), three simplexes  are shown for $m=3$ and various sets of the coefficients ${c_i}$. In Fig. (\ref{tt1}), the coefficients $c_i$ are approximately equal: $c_1=0.5$, $c_2=0.49$, and $c_3=0.51$. The seventh internal unstable fixed point is placed almost in the middle of the simplex, and the basins of attraction are almost of the same size. In Fig. (\ref{tt2}), the coefficients $c_i$ ($c_1=0.24$, $c_2=0.4$, $c_3=0.6$) are set as to assure the internal fixed point at the same position as the edge fixed point; hence these two fixed points, both unstable, collide. In Fig. (\ref{tt3}), the coefficients $c_i$ ($c_1=0.5$, $c_2=0.3$, $c_3=0.1$) are chosen as to make the seventh
fixed point out of the simplex. As we see, the pictures in Figs. (\ref{tt2}) and (\ref{tt3}) are qualitatively the same, except the order of the coefficients $c_i$. All the unstable fixed points are placed at the boundaries between the basins of attraction. \\

We conjecture that the same rules apply for higher dimensions of the system. Consider the case of a given $m$. Having fixed $m-3$ coordinates of a fixed point equal zero,
we are left with a three-dimensional system described above in this subsection. The same rule should apply to any dimensionality $m$ and $k$. This is a consequence of the model equations (Eq. \ref{de}): each subspace $W$ where $p_i=0$ for some subset of actors $i \in W$ is invariant, and the mere existence of these actors does not influence the system behavior.

%% ===========================================================================
\section{Discussion}
%% ===========================================================================

The structure of the fixed points, described above, allows to summarize the results as follows. Generic trajectories end up at one of the fixed points where $p_i=1$ for one player 
$i$, $p_j=0$ for all other $j$-s. Which one of such points is selected, depends on the set of the coefficients $c_i$ and on the initial values of ${p_i}$'s. The latter dependence
can be expressed in the form of basins of attraction of the stable fixed points. As a rule, the unstable fixed points are placed at the boundaries of the basins, hence they provide valuable information on these boundaries. Accordingly, for $m=2$ there are three fixed points, two stable (0,1) and (1,0), and third unstable at the edge between the stable ones. For $m=3$ there are three stable fixed points at the corners of the simplex, and three unstable fixed points at the edges of the triangle. Out of the coordinates of the latter, one is equal to zero. It is also possible that there is a seventh fixed point, either within the triangle surface or at the edge; in the latter case it coincides with one of fixed poins at the edge. This seventh point is also unstable. For higher $m$, a classification is possible along the same rules. For $m=4$ there are four stable fixed points (three coordinates of each equal to zero), and 6 unstable 'edge' fixed points with two coordinates equal to zero. Four further 'surface' fixed points can also appear, in the analogy to the case $m=3$. Finally, one unstable fixed point can appear within the volume of the simplex. If the latter happens to be at the surface, it coincides with the existing one at the same surface. In this way, the structure of all but the last of the fixed points for $m+1$-dimensional simplex can be reconstructed from the structure for $m$-dimensional one by adding coordinates equal zero to the existing fixed points.\\

There are some interesting analogies of these model results and the social reality. First is that basically, the winner is this player who engages minimally in the conflict. The winning strategy is to withdraw from the conflict, what can be carried out by setting $c_i=0$. If this strategy is accepted by all players, i.e. $c_i=0$ for each $i$, there is no conflict at all, and everybody stays with her/his initial power $p_i$. This is a kind of the Nash equilibrium \cite{straf}; whoever enters into conflict, loses. However, we know that people enter into conflict for various reasons, which are out of scope of the paradigm of rational players. Our results indicate that even the most aggressive player (with the largest value of $c_i$) can win, if his initial power $p_i$ is large enough. In this sense, the Matthew effect is reproduced here.\\

Summarizing, a model of conflict is proposed and explored, which takes into account the dependence of strategy on the actual power of a player. The structure of state space of model variables, controlled by nonlinear difference equations, reveals interesting phenomena as collisions of unstable fixed points. Further extensions of the model will include coalitions and state dependent strategies.

\vspace{1cm}
%% ===========================================================================
%\noindent
%{\bf Acknowledgements}\\
 % The work was partially supported by the PL-Grid Infrastructure.
%% ===========================================================================

{\bf References}\\

%%%%%%%%%%%%%%%%%%%%%%%%%%%%%%%%%%%%%%%%%%%%%%%%%%%%%%%%%%%%%%%%%%%%%%%%%%%%%%

\end{document}